\documentclass[aps,pre,twocolumn,superscriptaddress,showpacs,groupedaddress]{revtex4-1}
\usepackage{graphicx}
\usepackage{bm}
\usepackage{hyperref}
\usepackage[mathlines]{lineno}
\usepackage{amsmath}
\usepackage{amssymb}
\usepackage{mathtools}
\usepackage{times}

\usepackage[dvips]{color}
\usepackage{mathtools}
\usepackage{bm}

\begin{document}

\title{Small-angle scattering from three-phase systems: Investigation of the crossover between mass fractal regimes}

\author{Eugen M Anitas$^{1,2}$, Alexander Yu Cherny$^{1}$, Vladimir A Osipov$^{1}$ and Alexander I Kuklin$^{1,3}$}

\affiliation{$^{1}$Joint Institute for Nuclear Research, Dubna 141980, Moscow region, Russia}
\affiliation{$^{2}$Horia Hulubei National Institute of Physics and Nuclear Engineering, RO-077125 Bucharest, Romania}
\affiliation{$^{3}$Laboratory of Advanced Research on Membrane Proteins, MIPT, Moscow, Russia}
\email[]{anitas@theor.jinr.ru}

\date{\today}

\begin{abstract}
In this paper, we construct a three-phase model (that is, a system consisting of three homogeneous regions with various scattering length densities), which illustrate the behavior of small-angle scattering (SAS) scattering curves. Here two phases are a deterministic fractal embedded in another deterministic mass fractal, and they altogether are further embedded in a third phase, which can be a solution or solid matrix. We calculate SAS intensities, derive expressions for the crossover position (that is, the point where the power-law scattering exponent changes) as a function of control parameters, including size, concentration, and volumes of each phase. The corresponding SAS intensities from these models describe a succession of power-law regimes in momentum space where  both regimes correspond to mass fractals. The models can be applied to SAS data where the absolute value of the scattering exponent of the first power-law regime is higher than that of the subsequent second power-law regime, that is, the scattering curve of "convex" kind near the crossover position.
\end{abstract}

\maketitle
\section{Introduction}
Small-angle scattering (SAS)~\cite{svergun87:book,glatter82:book} experimental data (X-ray or neutron) often show a succession of power-law regimes, on a double logarithmic scale, whose scattering exponents are taking arbitrarily values from $-4$ to $-1$. Although this type of behavior is usually associated with a multilevel structure, the explanation of the crossover position is a more intricate problem and a method to extract the structural properties about each phase is lacking. Experimentally it has been shown that the crossover position in a multi-phase system can depend on the variation of the scattering length density (SLD) of the component phases~\cite{lebedev05} and a theoretical framework for describing these type of results has been developed in Ref.~\cite{cherny12} where this dependence of the crossover position was confirmed for the particular transition between the Porod and mass fractal~\cite{mandelbrot83:book} regimes. 

Recently, a practical formula which can be used to fit experimental SAS data from multi-phase systems has been derived~\cite{chernyArxiv}. It can be applied to SAS data which show a succession of power-law regimes: $I(q)\propto q^{\tau}$, where $\tau = D$ for mass fractals ($0<D<3$), and $\tau = 6-D$ for surface fractals ($2<D<3$)~\cite{martin87}. However, for fractal structures embedded in other fractal structures, the underlying physical structure of each phase and their spatial correlations are usually difficult to be clearly understood. In this paper we extend the results obtained in~\cite{cherny12} to the transition between two mass fractal regimes, we calculate the corresponding scattering intensity and estimate the crossover position. 

\section{{\label{sec:Models}}{Models}}

The three-phase model consists of a uniform matrix with scattering length density (SLD) $\rho_{0}$ (region 3) which contains mass fractal regions with SLD $\rho_{1}$ (region 1) which absorb other mass fractals with SLD $\rho_{2}$ (region 2). A similar model in which a mass fractal is embedded in a ball has been developed in~\cite{cherny12}. In Ref.~\cite{chernyArxiv} we relate this model to the \emph{type} $I$ structure ``fractal inside fractal".

The construction process of the fractal used here is very similar to that of the generalized Cantor and Vicseck fractals~\cite{chernyJACR10,chernyPRE11}. For the both fractals we start with a cube of edge $l_{0}$. We choose a Cartesian system of coordinates with the origin in the cube center and the axes being parallel to the cube edges. The iteration rule is to replace the cube of edge $l_{0}$ with twenty one smaller cubes of edge length $l_{1}=l_{0}/3$. The positions of the cubes with edge $l_{1}$ correspond to a Menger sponge structure with an additional cube in the center.  At the $m$th iteration the edge of the resulting cubes is $l_{m}=l_{0}/3^{m}$. 

\begin{figure}
\centering
\includegraphics[width=0.7\columnwidth,clip=true]{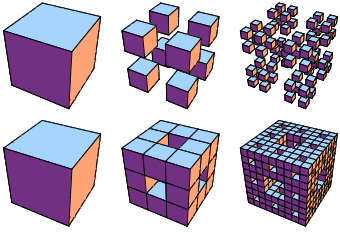}
\caption{(Color online) The initiator and the first two iterations of the fractals composing region 2 (upper panel: Cantor fractal) and respectively, region 1 (lower panel: Menger-like fractal). The three-phase system consists of region 1 absorbing region 2. }
\label{fig:fig1}
\end{figure}

The normalized scattering amplitude (form factor) of an object is $F(\bm{q}) = 1/V \int_{\mathrm{V}}{e^{-i\bm{q} \cdot \bm{r}}}\mathrm{d}\bm{r}$, where $V$ is the volume of the object and $F(0)=1$. Then for regions 1 and 2, the normalized form factors of the fractals at the $m$th iteration can be written as~\cite{chernyJSI,chernyJACR10,chernyPRE11} $F_{\mathrm{1,2}}^{(m)}(\bm{q})=F_{0}\left(\bm{q}l_{0}/3^{m}\right)\prod_{i=0}^{m}G_{\mathrm{1,2}i}(\bm{q})$
where $F_{0}(\bm{q})$ is the form factor of a cube of unit edge length, and $G_{\mathrm{1,2}m}$ are the generative functions of the fractals with the property that $G_{\mathrm{1,2}0}(\bm{q}) \equiv 1$. They are determined by the positions of the centers of the cubes for each iteration, and for the region 1 (Fig.~\ref{fig:fig1}, lower panel) it is given by
\begin{equation}
\label{eq:gfmengerlike}
G_{\mathrm{1}m}(\bm{q}) =\left(1+4\Gamma_{m}(\bm{q})+8\Lambda_{m}(\bm{q})\right)/21,
\end{equation}
where we denote $\Gamma_{m}(\bm{q}) = \cos(l_{m}q_{x})\cos(l_{m}q_{y})+\cos(l_{m}q_{x})\cos(l_{m}q_{z})+\cos(l_{m}q_{y})\cos(l_{m}q_{z})$ and $\Lambda_{m}(\bm{q}) = \cos(l_{m}q_{x})\cos(l_{m}q_{y})\cos(l_{m}q_{z})$.
For the region 2 we consider as generator the same cube with edge $l_{0}$ and we choose the generative function corresponding to a Cantor set (Fig.~\ref{fig:fig1}, upper panel) and we have~\cite{chernyJACR10}
\begin{equation}
\label{eq:gfcantor}
G_{\mathrm{2}m}(\bm{q}) = \cos(l_{m}q_{x})\cos(l_{m}q_{y})\cos(l_{m}q_{z}).
\end{equation}
Therefore, we shall observe the transition in the SAS intensity from the region 1, which is a Menger sponge-like fractal, to the region 2 (Cantor set): $q^{-D_{\mathrm{1}}} \rightarrow q^{-D_{\mathrm{2}}}$, where $D_{\mathrm{1}} = 2.77$ and $D_{\mathrm{2}} = 1.89$ are the corresponding fractal dimensions.  In order to assure a complete inclusion of the region 2 into region 1, the iteration number of region $2$ shall be higher than the iteration number of region 1.  

\section{{\label{eq:intensity}}{Scattering intensity}}

The scattered intensity for the multiphase system can be written as~\cite{cherny12,chernyArxiv}
\begin{equation}
I(q)=I(0)\left\langle \left|\alpha F_{\mathrm{1}}^{(m)}(\bm{q})+(1-\alpha)F_{\mathrm{2}}^{(p)}(\bm{q})\right|^{2}\right\rangle,
\label{eq:intensityfinal}
\end{equation}
where $I(0)=n V_{\mathrm{r}}^2(\rho_{1}-\rho_{0})^{2}\alpha^{-2}$ and $\alpha$ is the contrast parameter given by
\begin{equation}
\alpha=\left( 1+\frac{\rho_{2}-\rho_{1}}{\rho_{1}-\rho_{0}}\frac{V_{\mathrm{f}}}{V_{\mathrm{r}}}\right )^{-1}.
\label{eq:alpha}
\end{equation}

A more accurate description of SAS from a physical system should involve size polydispersity of the scatterers. Here we consider an ensemble of the three-phase fractal with different sizes taken at random and distributed according to a log-normal distribution function with relative variance $\sigma_\mathrm{r}$~\cite{chernyPRE11}. Thus, the average in Eq.~(\ref{eq:intensityfinal}) is taken both over angles and sizes. 

\section{Results and Discussions} 

\begin{figure}
\centering
\includegraphics[width=0.9\columnwidth,clip=true]{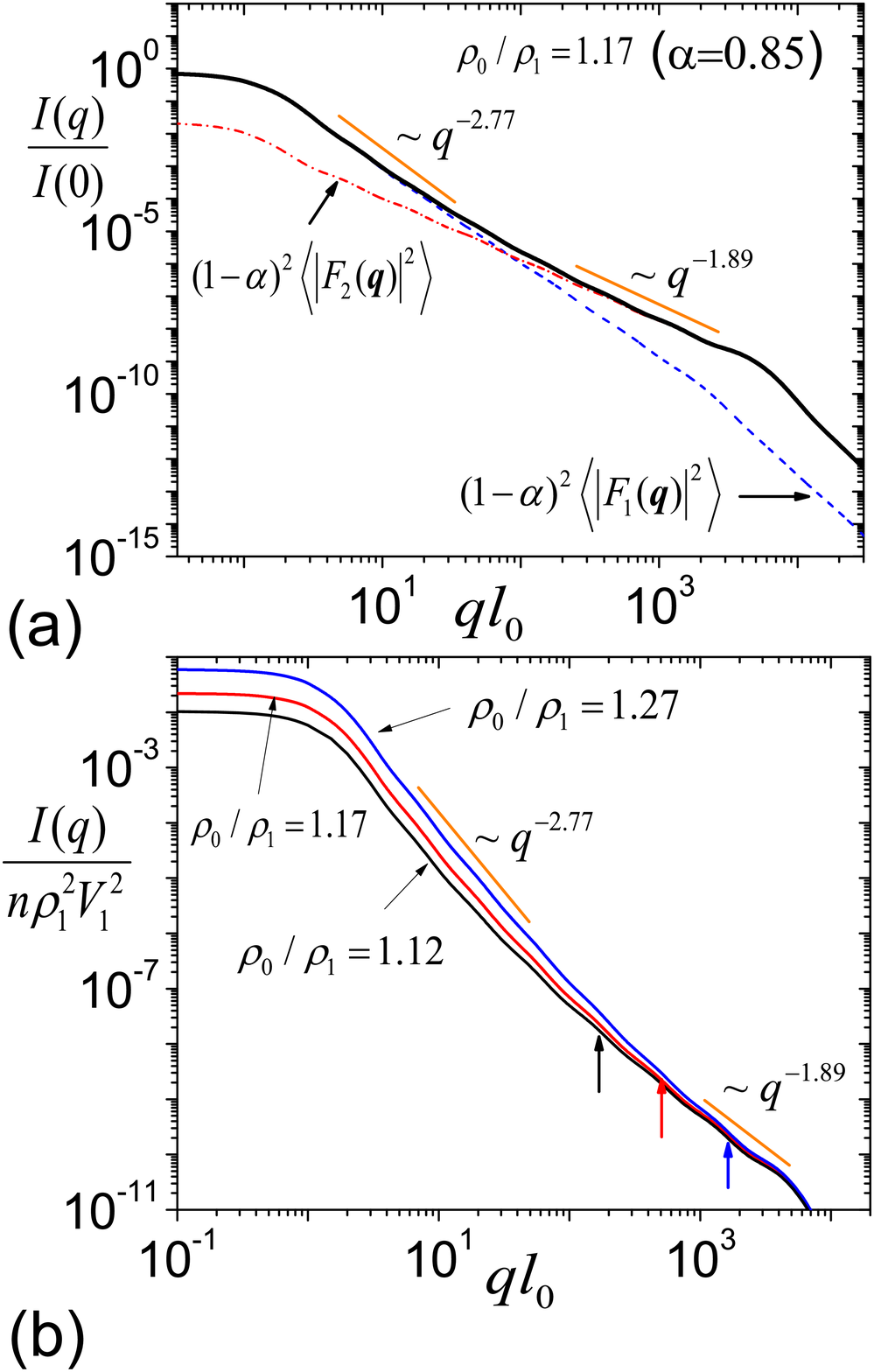}
\caption{(Color online) The scattering intensities from the three-phase model, where region 1 (Menger-like mass fractal) absorbs region 2 (Cantor mass fractal). The parameters are described in the text. (a) The scattering intensity (black: solid line) and the contributions of the region 2 (blue: dashed) and region 1 (red: dashed-dot) fractals. (b) Contrast variation of the scattering intensity. The vertical arrows represent the crossover position as estimated from Eq.~(\ref{eq:crossovermf}). Below the crossover point, the contribution of region 2 is negligible, while above it, the contribution of region 1 is negligible.}
\label{fig:fig2}
\end{figure}

The numerical results for the scattering intensity using Eq.~(\ref{eq:intensityfinal}), but neglecting the correlations (which can play a significant role only at low $q$~\cite{chernyArxiv}), are shown in Fig.~\ref{fig:fig2} for different values of the contrast parameter $\alpha$. One can clearly see the appearance of the crossover and we can estimate its position by a simple approximation in the limit of strongly developed polydispersity~\cite{chernyArxiv}
\begin{equation}
\langle|F_{\mathrm{1,2}}^{(m)}(\bm{q})|^{2}\rangle \propto 
\begin{dcases}
   1, &q \lesssim {2\pi}/{l_{0}},\\
   {(ql_{0}/2\pi)^{-D_{\mathrm{1,2}}}}, & {2\pi}/{l_{0}} \lesssim q \lesssim {2\pi}/{l},\\
   \frac{1}{u}(ql/2\pi)^{-4}, &q \gtrsim {2\pi}/{l},\end{dcases}
\label{eq:partialintensitiessurr}
\end{equation}
where $u$ is of the order of $(l_{0}/l)^{D_{\mathrm{1,2}}}$.

For a given $q$, the amplitude of region 1, $F_{\mathrm{1}}^{(m)}(\bm{q})$, and the amplitude of region 2, $F_{\mathrm{2}}^{(m)}(\bm{q})$, in the r.h.s. of Eq.~(\ref{eq:intensityfinal}) can be of \emph{different orders of magnitude}, provided the \emph{corresponding exponents are not equal}. Then only one of the amplitudes dominates. Therefore in a good approximation, the crossover position can be found by equating the corresponding intensities $\alpha^2 \langle|F_{\mathrm{1}}^{(m)} (\bm{q}_{\mathrm{c}})|^{2}\rangle \simeq (1-\alpha)^2 \langle|F_{\mathrm{2}}^{(p)} (\bm{q}_{\mathrm{c}})|^{2}\rangle$, where one can use the estimations from Eq.~(\ref{eq:partialintensitiessurr}). This gives us the following expression for the crossover position
\begin{equation}
q_{\mathrm{c}} \simeq \frac{2\pi}{l_0}\left( \frac{(1-\alpha)^2}{\alpha^2} \right)^{1/(D_{\mathrm{2}}-D_{\mathrm{1}})}.
\label{eq:crossovermf}
\end{equation}

One can clearly observe the Guinier region (which corresponds to both region 1 and region 2), fractal regions of the region 1 and region 2, and a Porod region (Fig.~\ref{fig:fig2}), where we choose $\rho_{2}/\rho_{1}=0.9$ for all the plots, $\sigma_{\mathrm{r}}=0.4$ for the relative dispersion, and fractal iteration number $m_1=6$, and respectively $m_2=7$. Then, the values of $\alpha$ depend on the  ratio $\rho_{0}/\rho_{1}$. For arbitrarily values of $\alpha$ one obtains a direct and \textit{smooth} transition between fractal regions. The crossover positions are estimated from Eq.~(\ref{eq:crossovermf}) and their variation is depicted in Fig.~\ref{fig:fig2}b.

Choosing the SLDs so that the contrast $\rho_{2}-\rho_{1}$ between the region 1 and region 2 is suppressed or setting the volume of region 2 small enough, $\alpha$ gets closer to one and as a consequence, a part of the curve corresponding to region 2 is ``absorbed" by the curve corresponding to region 1. Therefore, at sufficiently small values of $|1-\alpha|$, the fractal region of the region 2 may not be observed at all (when $\alpha \gtrsim 0.9$ at the chosen values of parameters). 

\section{Conclusions}

We develop a theoretical model which describes SAS from multi-phase fractal systems, where one mass fractal (one phase) is embedded into another mass fractal (a second phase) and altogether the combined system is further put into a solution or a solid matrix (a third phase). We derive an analytical expression for the SAS intensity from a mixture of such multi-phase systems. The scattering intensity is characterized by the presence of two successive mass fractal regimes with various scattering exponents and by the appearance of a crossover between these power-law regimes. We show that the crossover position is controlled by the effective contrast parameter which in turn, depends on the relative values of the SLD of each phase and their volumes, and estimate its position.

The developed model can be extended to describe scattering from various complex structures embedded into each other, such as mass fractal into a surface fractal, surface into a surface or surface into a mass fractal.

\section*{Acknowledgments}
We acknowledge financial support from JINR~–~-~IFIN~-~HH projects and from JINR grant No. 13-302-02.

\section*{References}

\bibliography{sasMM}

\end{document}